# Statistical Modelling of Computer Network Traffic Event Times

Matthew Price-Williams
Department of Mathematics, Imperial College London
and
Nick Heard
Department of Mathematics, Imperial College London
Heilbronn Institute for Mathematical Research, University of Bristol

December 2017

## Abstract

This paper introduces a statistical model for the arrival times of connection events in a computer network. Edges between nodes in a network can be interpreted and modelled as point processes where events in the process indicate information being sent along that edge. A model of normal behaviour can be constructed for each edge in the network by identifying key network user features such as seasonality and self-exciting behaviour, where events typically arise in bursts at particular times of day. When monitoring the network in real time, unusual patterns of activity could indicate the presence of a malicious actor. Four different models for self-exciting behaviour are introduced and compared using data collected from the Imperial College and Los Alamos National Laboratory computer networks.

*Keywords:* Computer network. Hawkes process. Wold process. Changepoint detection.

# 1 Introduction

Statistical anomaly detection tools (Lazarevic et al. 2003, Neil et al. 2013) have an important role to play in the next generation of cyber-security defences, complementing more traditional "signature-based" techniques which rely on packet inspection to match known indicators of malicious content held in a database. In contrast, anomaly detection techniques harvest data on *normal* behaviour in a computer network, and monitor for any significant deviations in observed traffic from probability models built on those historic data. An advantage of anomaly detection techniques is that attack vectors which have not previously been observed may still be detected, and for this reason any departures from normality are of potential interest.

We construct a model of a computer's normal behaviour in an enterprise computer network using only the observed event times of computer-computer connections. Lambert et al. (2001) modelled timing patterns in network data using a "dynamic Poisson timing model" with time partitioned into seasonal periods and an exponentially weighted moving average intensity parameter. The approach considered in this article provides a more general counting process framework that captures the self-exciting nature of computer network flow data, where event data can be seen to occur in bursts. This is partly due to human activity, with users having sessions of activity connecting from a particular client to a particular server, and partly due to the data collection process where a stream of packets are aggregated into grouped summaries of packet flows with some arbitrariness in how the records are divided.

Let $y_1, y_2, \ldots$ be the observed event times in seconds of a counting process $Y(t)$ of con-nections between two computer nodes. The conditional intensity of $Y(t)$ will be assumed to have a general form

$$\lambda_Y(t) = \mu(t)\left(\lambda + \sum_{i:y_i<t}\psi_i\left(\int_{s=y_i}^{t}\mu(s)\right)\right) = \mu(t)\left(\lambda + \sum_{i:y_i<t}\psi_i\{M(t)-M(y_i)\}\right) \quad (1)$$

where $\mu(t) \geq 0$ is a function which captures seasonal variation in the intensity and

$$M(t) = \int_{s=0}^{t}\mu(s). \quad (2)$$

For a fixed, non-increasing function $\omega(u) \geq 0$, $\psi_i(u)$ takes one of two forms:

Hawkes process: $\quad \psi_i(u) = \omega(u)$

Wold process: $\quad \psi_i(u) = \omega(u) - \sum_{i'<i}\psi_{i'}(u)$

The process models of Hawkes (1971) and Wold (1948) fundamentally differ in the following way: Under the Hawkes model, all events occurring before time $t$ potentially contribute to the intensity at $t$; in contrast, defining $y_t^* = \max_i\{y_i : y_i < t\}$, the Wold process intensity model simplifies to

$$\lambda_Y(t) = \mu(t)\left(\lambda + \omega\{M(t)-M(y_t^*)\}\right)$$

and so only depends on the time since the most recent event. Both provide potentially plausible models for the burstiness in computer network data, and so their performance will be compared.

The compensator for $Y(t)$ is

$$\Lambda_Y(t) = \int_{s=0}^{t} \lambda_Y(s)ds = \lambda M(t) + \sum_{i:y_i<t} \Psi_i\{M(t) - M(y_i)\}, \qquad (3)$$

where $\Psi_i(t) = \int_{s=0}^{t} \psi_i(s)ds$.

The structure of the remainder of the article is as follows: Section 2 describes the motivating network traffic flow data, obtained from the computer networks of Imperial College and Los Alamos National Laboratory. Section 3 then introduces a piecewise constant model for the inherent seasonalities present in computer network traffic. Section 4 presents four models for self-exciting behaviour using Hawkes and Wold processes. Section 5 describes how changepoint detection algorithms can be used to estimate the parameters of both the seasonal and self-exciting components of the models. The performance of these approaches to modelling computer network traffic are compared in Section 6. Section 7 presents an alternative model for discrete event time data, and demonstrates performance over an entire computer network.

## 2 Computer network data

Two sources of computer network traffic data are considered: network flow ("NetFlow") data from the Imperial College computer network, and authentication logs obtained from the enterprise network of Los Alamos Laboratory (LANL)(Turcotte et al. 2017). NetFlow records are a high level aggregation of the packets sent between two IP addresses over the same ports, under the same protocol in a short space of time, and summarise what might be interpreted as a single communication between the two addresses. NetFlow data are difficult to anonymise, and so cannot be publicly shared.

The authentication log data have been anonymised and are publicly available (Turcotte et al. 2017), and specify the times at which users performs authentication events on computers within the internal network. An event type is recorded that indicates which kind of authentication event occurred, such as a network log-on, interactive log-on, a workstation screen lock, and so on. Turcotte et al. (2017) and Price-Williams et al. (2017) noted that some authentication event types exhibit strong periodic patterns, indicating automated network activity. For this analysis, we only consider event types that are considered "interactive" and can be linked to a user being present at their computer.

Figure 1 plots an example of both data sources. In the left image NetFlow event start times, which are recorded in milliseconds, are plotted for connections from one Imperial College IP address to a particular internet sever over a 4-week period. The right image plots the "interactive log-on" events, which were recorded in seconds, for a random user (user 689229) from the LANL network. It is apparent that the event times in both examples exhibit seasonal variations, with all events occurring within the typical hours of a working day, and clear two day breaks that correspond to weekends. More interestingly, particularly in the NetFlow data, the events are seen to occur in bursts, even when diurnal patterns have been taken into account.

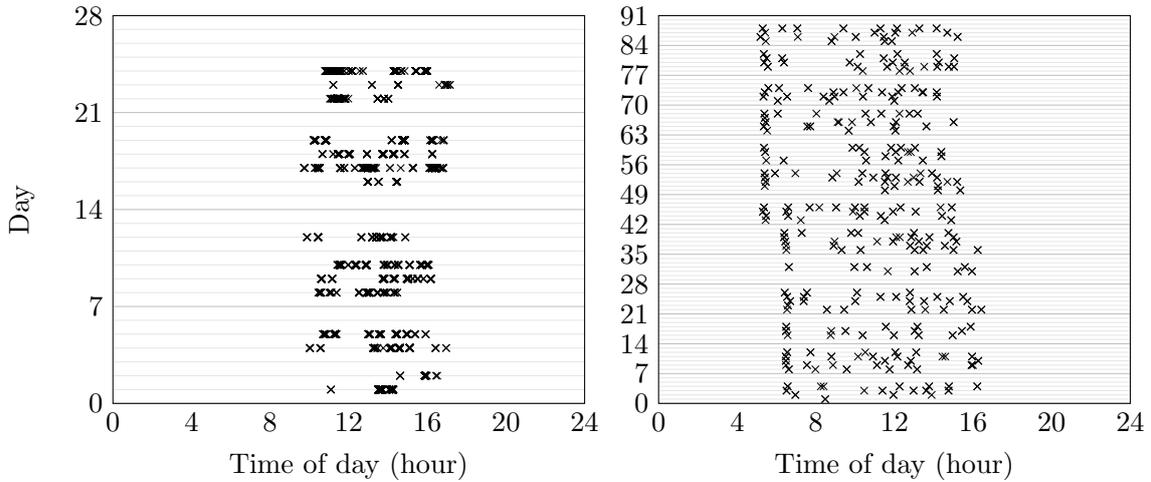

Figure 1: Examples of user-driven computer network event data. Left: Start times of all connections between one Imperial College IP address and a particular internet server, recorded over 28 days. Right: Times of all interactive log-on events made by user 689229 from the LANL network, over 90 days.

## 3 Modelling seasonality

Seasonal patterns appear in computer network traffic data since human users are more likely to be active on weekdays during the day time than they are at night time or at the weekend. In the counting process conditional intensity (1), these seasonal variations are represented by the function $\mu(t)$. For seasons of total duration $S$, models for $\mu(t)$ can be assumed to be repeated in blocks of time $S$, such that $\mu(t) \equiv \mu(t \bmod S)$ for all $t > 0$. Assuming a weekly seasonality of $S = 604{,}800$ seconds, we therefore simply require a model for $\mu(t)$ on the time-domain $[0, S]$.

Within a single week there will be daily fluctuations in activity, with weekend days in particular likely to witness fewer network events than weekdays. However, besides these fluctuations in volume, it will be further assumed that the activity patterns each day will have the shape. Formally, we assume

$$\mu(t) = \tilde{\mu}(t \bmod S')\bar{\mu}_{\lfloor (t/S' \bmod 7 \rfloor}$$

where $S' = S/7$ is the duration of one day in seconds, $\lfloor x \rfloor$ is the integer part of $x$, $\tilde{\mu}(t)$ is a probability density function on $[0, S']$ and $\bar{\mu}_0, \ldots, \bar{\mu}_6$ are positive daily multipliers such that the seasonal component of the conditional intensity (1) on the $i$th day of the week at the time of day $t$ seconds is $\tilde{\mu}(t)\bar{\mu}_i$.

Given some historical event data, the parameters $\bar{\mu}_0, \ldots, \bar{\mu}_6$ can be simply estimated by the average number of events occurring on each week day, and so it simply remains to estimate the density function $\tilde{\mu}(t)$ of event arrivals throughout the day. For computational tractability and simplicity, $\tilde{\mu}(t)$ will be assumed to be piecewise constant with an unknown number of changepoints $m$, denoted $\sigma_1, \ldots, \sigma_m$, ordered such that $0 = \sigma_0 < \sigma_1 < \ldots < \sigma_m < \sigma_{m+1} = S'$; let $\tilde{\mu}_0, \tilde{\mu}_1, \ldots, \tilde{\mu}_m$ be the corresponding densities in each changepoint segment, such that

$$\tilde{\mu}(t) = \sum_{i=0}^{m} \tilde{\mu}_i \mathbb{I}_{[\sigma_i, \sigma_{i+1})}(t) \tag{4}$$

and $\sum_{i=0}^{m} \tilde{\mu}_i(\sigma_{i+1} - \sigma_i) = 1$. Estimation of (4) will be considered later in Section 5.

## 4 Modelling self-exciting behaviour

In the general counting process conditional intensity (1), self-exciting behaviour is represented by the function $\psi(u)$ and the scalar $\lambda$. Given an estimate for $\mu(t)$, modelling self-exciting behaviour can be simplified using the time-rescaled process

$$Z(t) = Y(M^{-1}(t)), \qquad (5)$$

where $M^{-1}$ is the inverse of (2). For each event time $y_i$ in $Y(t)$, let $z_i = M(y_i)$ be the corresponding event time in $Z(t)$. The intensity of $Z(t)$ is then given by

$$\lambda_Z(t) = \lambda + \sum_{i:z_i<t} \psi_i\left(t - z_i\right),$$

which further simplifies to

$$\lambda + \sum_{z_i<t} \omega(t - z_i) \qquad (6)$$

for the Hawkes model and

$$\lambda + \omega\left(t - z_t^*\right) \qquad (7)$$

for the Wold model, where $z_t^* = \max_i\{z_i : z_i < t\}$.

The choice of the non-increasing excitation function $\omega$ controls the increase in a user's intensity following each event and the subsequent rate of decay. Two alternatives are con-

sidered:

$$\text{Exponential:} \quad w(u) = \alpha \exp\left(-\beta(u)\right), \tag{8}$$

$$\text{Piecewise constant:} \quad w(u) = \sum_{i=0}^{\ell} \lambda_i \mathbb{I}_{[\tau_i, \tau_{i+1})}(u), \tag{9}$$

where $\alpha, \beta > 0$ in (8), and (9) is a step function with $\ell$ changepoints $0 \equiv \tau_0 < \tau_1 < \ldots < \tau_\ell$ and decreasing step heights $\lambda_0 > \ldots > \lambda_{\ell-1} > \lambda_\ell \equiv 0$. Substituting either of the two excitation functions (8) and (9) into the Hawkes and Wold models (6) and (7) gives rise to four different models for self-exciting behaviour. Cartoon illustrations of these models are shown in Figure 2.

## 4.1 Parameter estimation

Given the event times $z_1, z_2, \ldots, z_n$ of the time-rescaled counting process (5) observed on $[0, T]$, from Daley & Vere-Jones (2007) the likelihood function of a model for $\lambda_Z(t)$ is given by

$$\exp\left(-\int_0^T \lambda_Z(u) du\right) \prod_{i=1}^{n} \lambda_Z(z_i). \tag{10}$$

Numerical maximum likelihood estimation is straightforward for the exponential excitation function (8), since in this case there are just two unknown parameters under either the Hawkes or Wold models. More details are provided in Ozaki (1979), Laub et al. (2015). In contrast, the non-increasing step function excitation model (9) has an unknown number of changepoint parameters to optimise. Under the simpler Wold model, the number of changepoints and their locations can be successfully estimated using a changepoint algorithm, using the Bayesian information criterion (BIC) (Schwarz 1978) to prevent over-fitting the number of parameters. This will be considered in detail in Section 5.

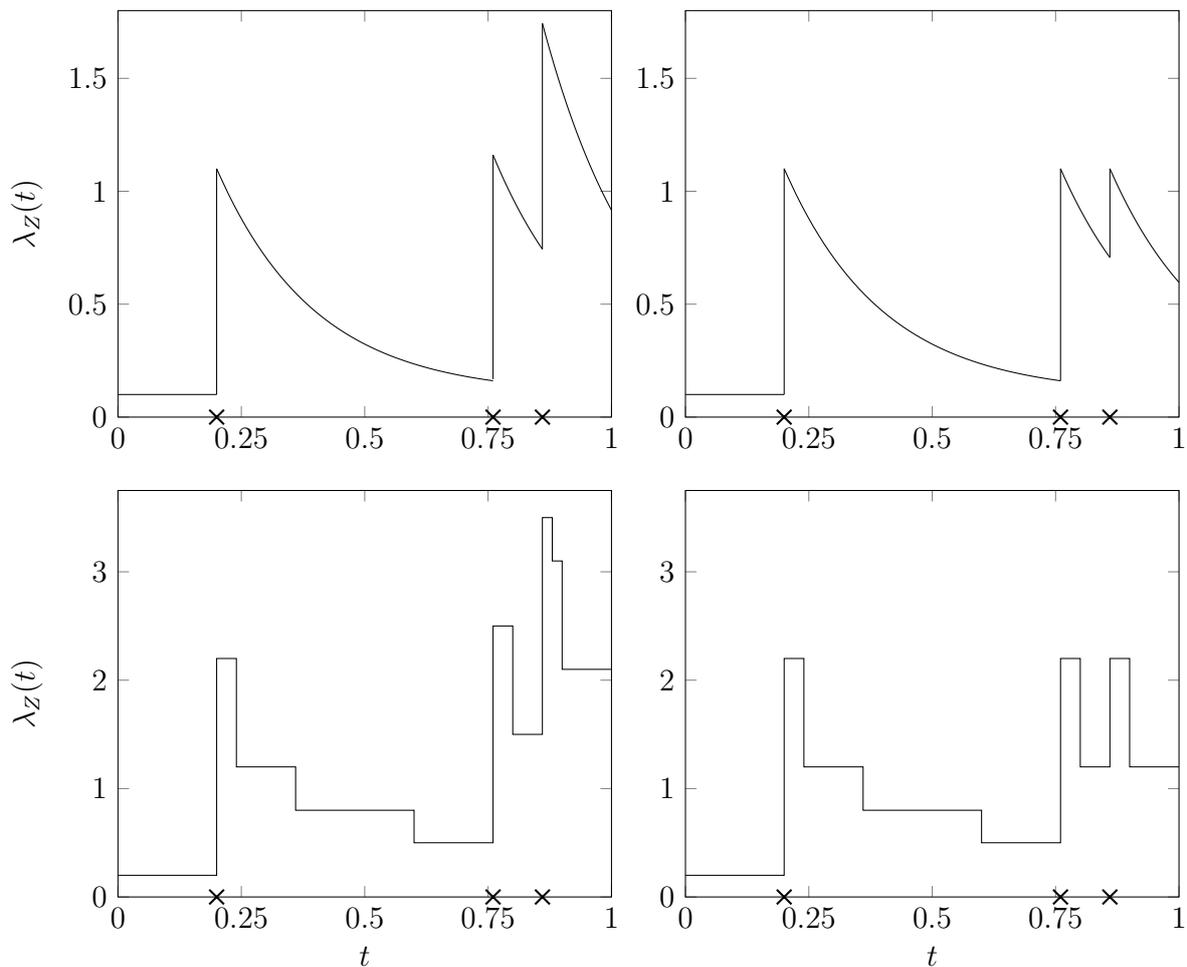

Figure 2: Examples of realised conditional intensity functions for a point process with three events. The left column corresponds to intensities estimated using the Hawkes model (6), and the right column corresponds to the Wold model (7). The excitation function $\omega$ is either an exponential decay (top row) or a step function (bottom row).

Under the full Hawkes model, even applying changepoint approaches is too complex. Instead, a simpler version of the Hawkes-step function model will be considered, where the number of steps $\ell$ is fixed to be 1. In this case, simple numerical optimisation of (10) can be deployed.

## 5 Changepoint detection

Estimation of the changepoints for both the seasonal density (4) and for the piecewise constant excitation function (9) under the Wold model can be efficiently performed using pruned exact linear time (PELT) method of Killick et al. (2012). For an observed sequence of random variables $x_{1:n}$, the algorithm finds changepoints $0 \equiv \tau_0 < \tau_1 < \ldots < \tau_m < \tau_{m+1} \equiv n$ which minimise the Bayesian Information Criterion (Schwarz 1978),

$$\sum_{i=1}^{m+1} -2\log\left(\mathcal{L}(x_{\tau_{i-1}+1:\tau_i})\right) + \alpha \log n,$$

where $\mathcal{L}$ is an estimated likelihood function and $\alpha > 0$ notionally represents the number of additional free parameters introduced to the model by adding a changepoint. When $x_{1:n}$ are the event times from an inhomogeneous Poisson process,

$$\mathcal{L}(x_{\tau_{i-1}+1:\tau_i}) = \exp(\{-(\tau_i - \tau_{i-1})\})(\tau_i - \tau_{i-1})/(x_{\tau_i} - x_{\tau_{i-1}})$$

and $\alpha = 2$, since introducing a new changepoint adds two parameters: the location of the changepoint, and the intensity within the new segment.

To estimate the seasonal density, historical event times from $Y(t)$ are mapped onto $[0, S']$ and changepoint detection for (4) proceeds with PELT by treating these as a Poisson process.

For estimating the excitation step function (9) in the Wold model, the event times $z_{1:n}$ from the transformed process $Z(t)$ (5) must be further transformed into Poisson process event times before the same estimation procedure can be used. Let $d_i = z_{i+1} - z_i$, $i = 1, \ldots, n-1$ be the waiting time between successive events and let $d_{(1)}, \ldots, d_{(n-1)}$ be the corresponding order statistics. Defining $\delta_i = \sum_{j=1}^{i}(n - j + 1)d_{(n-j)}$, then under the excitation model (9), $\delta_1, \ldots, \delta_{n-1}$ are event times from an inhomogeneous Poisson process with conditional intensity at $t$ given by

$$\lambda + \sum_{i=0}^{\ell} \lambda_i \mathbb{I}_{[\tau_i, \tau_{i+1})}(t). \quad (11)$$

Let $\hat{\tau}_{1:\hat{\ell}}$, $\hat{\lambda}$ and $\hat{\lambda}_{1:\hat{\ell}+1}$ be the maximum likelihood changepoints and intensities subject to the constraint that $\hat{\lambda}_{1:\hat{\ell}+1}$ are non-increasing. These can be calculated recursively,

$$\hat{\tau}_j = \underset{i \in \hat{\tau}_{j-1}+1, \ldots n-1}{\arg\max} \left( \frac{i - \hat{\tau}_{j-1}}{\delta_i - \delta_{\hat{\tau}_{j-1}}} \right), \quad \hat{\lambda}_j = \hat{\lambda} - \left( \frac{\hat{\tau}_j - \hat{\tau}_{j-1}}{\delta_{\hat{\tau}_j} - \delta_{\hat{\tau}_{j-1}}} \right).$$

To ensure the BIC estimate of the intensity (11) is still non-increasing, the PELT algorithm is constrained here to only allow changepoints which are a subsequence from the full, constrained maximum likelihood solution $\hat{\tau}_{1:\hat{\ell}}$.

# 6 Modelling NetFlow event times

The conditional intensity function (1), assuming either the Hawkes (6) or Wold (7) models, with exponential (8) or piecewise constant (9) excitation functions are now fit to the NetFlow event times from Figure 1, representing the times an Imperial College computer contacted an internet server over a 28-day period. Under each model, the parameters are estimated using the first two weeks of data, and then goodness-of-fit is compared on the second two weeks of data.

To examine the effect of modelling seasonality, it is first assumed that $\psi_i(u) = 0$ in (1), and that the seasonal density $\tilde{\mu}(t)$ is given either by (4) or else it is constant and equal to $1/S$. The maximum likelihood estimate for the parameter $\lambda$ is given simply by $n/T$, the average number of events observed per unit of time.

If the model for the conditional intensity (1) is correct, then the time rescaling theorem (Brown et al. 2002) states that $\Lambda(y_1), \ldots, \Lambda(y_n)$ are the event times of a homogeneous Poisson process with unit rate, where $\Lambda(t)$ is the compensator function (3). The inter-arrival times of this time-transformed process, $\Lambda(y_i) - \Lambda(y_{i-1})$, $i = 1, \ldots, n$ with $y_0 \equiv 0$, should therefore be exponentially distributed with rate 1. Under the null hypothesis of the intensity model being correct, the upper tail $p$-value is given by

$$p_i = \exp\left(-(\Lambda(y_{i+1}) - \Lambda(y_i))\right). \tag{12}$$

Under the intensity model, $p_i \sim \text{Uniform}(0, 1)$, while large (small) $p$-values suggest smaller (larger) than expected waiting times between event. A KolmogorovSmirnov (KS) test (Massey Jr 1951) can be used to assess the goodness of fit of the event times to the model by comparing the empirical cumulative distribution function (CDF) of $p_{1:n}$ to the CDF of a Uniform$(0, 1)$ distribution. Table 1 shows the KS test statistic for each of the proposed models for the conditional intensity (1). For a visual assessment, Q–Q plots of the $p$-values $p_{1:n}$ are presented in Figure 3.

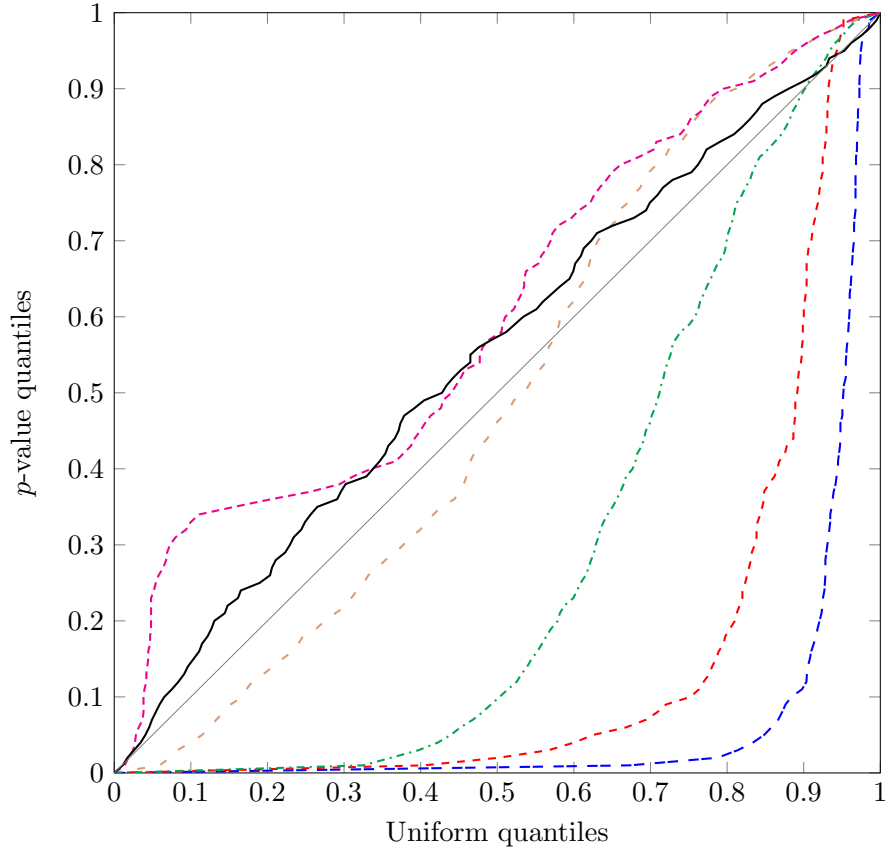

Figure 3: Q–Q plots for different conditional intensity models for connections from a client in the Imperial College computer network to an internet server, assuming the intensity is constant (-- -) or seasonally adjusted (- - -); a seasonally adjusted Hawkes model with exponential (- - ) or step function excitation (-·-·-); or a seasonally adjusted Wold model with exponential (----) or step function excitation (——).

The seasonal baseline model provides only a limited improvement over the assumption of a constant intensity, suggesting that capturing seasonal variability alone is not sufficient for realistic modelling of the arrival times of NetFlow events; the shape of the Q-Q curve for the seasonal model in Figure 3 implies the $p$-values are too large, meaning the majority of the events arrive more quickly than expected. Incorporating any of the four models for capturing self-excitation causes a significant jump in performance. The Wold step function model narrowly outperforms the Hawkes exponential method, possibly because the number of parameters is not fixed and is estimated from the data using a changepoint detection algorithm, providing greater flexibility.

| Conditional intensity model | KS test statistic |
|---|---|
| Homogeneous | 0.799 |
| Seasonal | 0.668 |
| Seasonal + Wold exponential | 0.228 |
| Seasonal + Hawkes exponential | 0.137 |
| Seasonal + Wold step | 0.101 |
| Seasonal + Hawkes step | 0.415 |

Table 1: KS test statistics comparing the the goodness-of-fit of the empirical CDF of $p$-values (12) to the CDF of a Uniform$(0,1)$ distribution, for different models for the conditional intensity.

## 6.1 Modelling computer network data without seasonality

In some situations it may not be feasible to usefully model the seasonal component of a user's conditional intensity (1); for example, if a user has irregular work patterns. Furthermore, it is possible for the self-exciting component to largely compensate for a poorly specified seasonal model, such as $\mu(t) \equiv 1$, as the self-exciting terms will naturally increase the intensity function more during seasonal periods where underlying activity is higher.

Table 2 shows the KS test statistics for the same data used for Table 1, but when no longer modelling seasonality, and the corresponding Q–Q plots are shown in Figure 4. Performance is similar across the two tables, with some methods such as the Wold step function excitation model performing even better without attempting to model seasonality. On this evidence, the computational inconvenience of including a seasonal component to an intensity model does not seem justified.

| Conditional intensity model | KS test statistic |
|---|---|
| Wold exponential | 0.289 |
| Hawkes exponential | 0.149 |
| Wold step | 0.092 |
| Hawkes step | 0.267 |

Table 2: KS statistics for four models of self-exciting behaviour applied to the Imperial College NetFlow data.

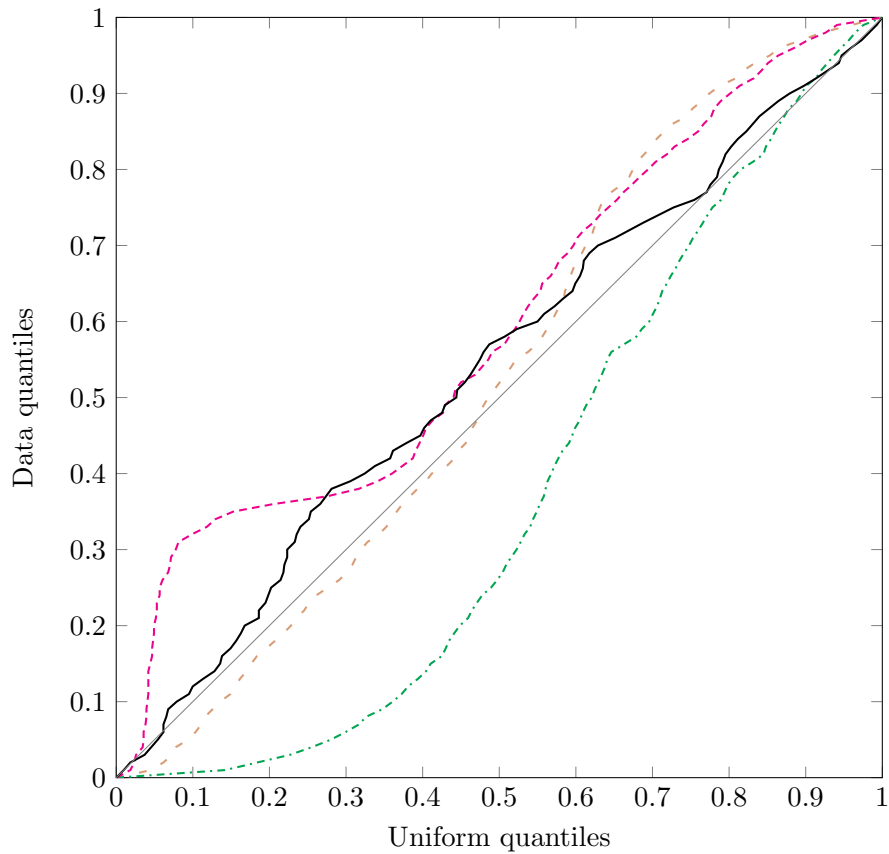

Figure 4: Q–Q plots for connections from a client in the Imperial College computer network to an internet server, assuming no seasonal adjustment. The conditional intensity is a Hawkes model with exponential (- -) or step function excitation (-·-·-), or a Wold model with exponential (- - -) or step function excitation (——).

# 7 Discrete event times

In some real world cyber-security data sets, event times may be recorded quite crudely. For example, the authentication logs from the Los Alamos National Laboratory enterprise network introduced in Section 2 have event times recorded only to the nearest second. Particularly when the event times are bursty, this can lead to several events appearing to be observed at the same moment in time, and so a Poisson process model is no longer appropriate. One solution could be to bin all events that appear in the same second and treat them as one combined event in any further analysis. A problem with this solution, particularly in anomaly detection, is that a bursty sequence of events may actually represent anomalous behaviour and binning the events within one second may weaken or remove the anomalous signal. An alternative approach considered here is to explicitly model the arrivals as a discrete time process.

Following on from Section 6, attention is restricted here to the best performing model from that section, which was the Wold model with self-excitation described by a step function, with no seasonal component. This model is particularly straightforward to convert into discrete time. In continuous time, under the Wold model for $Y(t)$ with step function self-excitation, the conditional intensity function simplifies to

$$\lambda_Y(t) = \lambda + \sum_{i=0}^{\ell} \lambda_i \mathbb{I}_{[\tau_i, \tau_{i+1})}(t - y^*(t)), \tag{13}$$

where $\lambda_1 > \lambda_2 > \ldots > \lambda_\ell > 0$. This translates to piecewise exponentially distributed waiting times between events. To recast this model in discrete time, first the changepoints $\tau_i$ can be constrained to be integer-valued; second, since the geometric distribution is the discrete analogue for the exponential distribution (both share the lack of memory property), it will be assumed that on entering segment $S_i = \{\tau_i, \ldots, \tau_{i+1} - 1\}$ without having observed an event, the next event occurs in $S_i$ with probability

$$1 - \exp\{-\lambda_i(\tau_{i+1} - \tau_i)\}, \tag{14}$$

and if so, the event time follows a truncated geometric distribution on $S_i$. This formulation allows an unbounded number of events to occur at each time point. The parameters of (13) are estimated using the same methodology as in Section 5, with the likelihood function adapted to the revised model (14).

## 7.1 Analysis of user authentication event data

As mentioned above, the authentication data from the Los Alamos National Laboratory (LANL) computer network, Section 2, are only recorded to the nearest second with many coincidental values, and so modelling must proceed in discrete time. For a comparison with the proposed approach, a baseline homogeneous model is also implemented with a discrete hazard function which is constant over time.

For each user in the network, the first 28 days of data are used to estimate the parameters of both models. The performances are then tested on the remaining 62 days of data. The analysis is restricted to the 3119 users in the LANL network with at least 200 events in both sections of data. For each user, the Kolmogorov-Smirnov (KS) test statistic is calculated to measure how well the two discrete models fit the test data.

Figure 5 displays box plots for each model of the distributions of the KS statistics across all of the users. The discrete-time Wold step function excitation model strongly outperforms the homogeneous model for most users in the network. However, the KS statistic for some users is still very large. This may be because the user's behaviour markedly changes at some point during the period of the test data. For a more robust procedure, a cyber analyst might wish to fit these models dynamically, so that changes in a user's behaviour could be identified and accounted for.

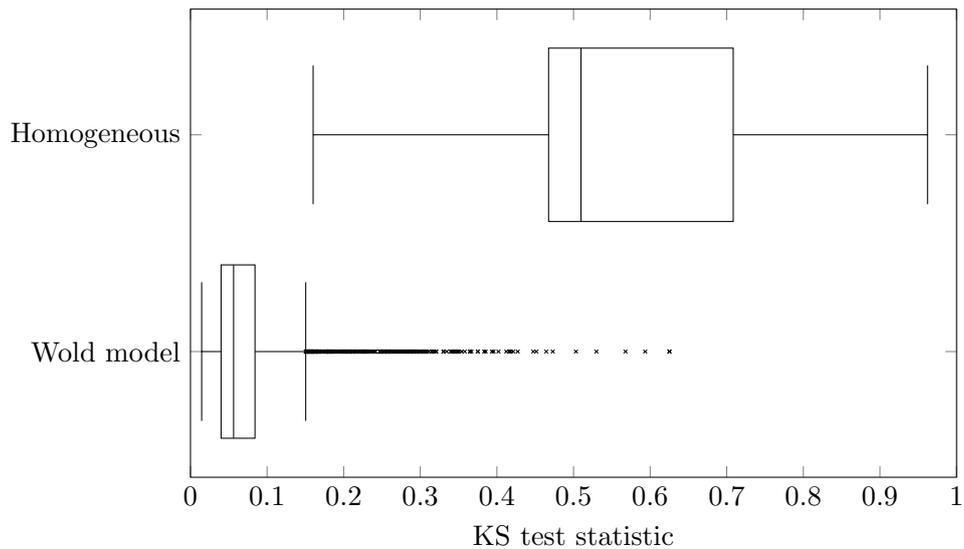

Figure 5: Box plots representing the distribution of KS test statistics for modelling user driven network behaviour, across 3119 users from the Los Alamos computer network, assuming either a time-homogeneous discrete hazard or the Wold step model.

# 8 Conclusion

A general framework for specifying the conditional intensity for normal, user-driven computer network behaviour has been presented, capturing key user behavioural features such as seasonality and self-exciting burstiness. Such models have important applications in anomaly detection in statistical cyber-security, (Neil et al. 2013, Turcotte 2013), potentially decreasing the number of false positive detections through more realistic models of normal network and user behaviour.

A Wold model for self-excitement, using step functions to model decay in excitation, was shown to perform very well across a large sample of users, and negated the need for modelling any seasonal variations in the data, such as diurnal activity patterns. The Hawkes exponential and Hawkes step function models could be used to detect anomalies along each edge of a computer network by monitoring the peak magnitude of the estimated conditional intensity; if this intensity rose above a nominated threshold value, a flag could be recorded indicating a possible attack with unusually high activity.

# 9 Supplementary Material

Supplementary materials avaliable online at `https://github.com/Matt0312/SToCND` contain python code to implement the methods introduced in the article.